\MAINTITLE={ Gas in Shearing Density Waves }
\SUBTITLE={ ????? }
\AUTHOR={Markus Demleitner}
\OFFPRINTS={ ????? }
\INSTITUTE={Astronomisches Rechen-Institut, M\"onchhofstra\ss e
12-14, 69120 Heidelberg, Germany}
\DATE={Received \dots, accepted \dots}
\ABSTRACT={
We examine the development of a transient spiral arm in a disk galaxy
made up of both gas and stars. To this end we have performed numerical
simulations in a shearing sheet (basically a rectangular patch of a
disc) that contains gas in the form of clouds behaving like Brahic's
(1977) sticky particles, and stars that appear as a background continuum
providing the perturbation forces. These are computed from the theory of
swing amplification, using Fuchs' (1991) work.  We describe the evolution
of our model under a single and under recurring swing amplification
events, discerning three phases.  Furthermore, we give an interpretation
of this evolution in terms of a variation of the epicyclic frequency
with the distance to the wave crest.  We also assess the importance of
self gravity in the gas for our results.

}
\KEYWORDS={Galaxies: kinematics and dynamics -- ISM: kinematics and
dynamics}
\THESAURUS={09.11.1 -- 11.11.1}
\maketitle

\newif\ifshowmods
\newif\ifproofmaking
\proofmakingfalse
\showmodsfalse

\def\I{{\rm i}}

\def\kxin{k_x^{\rm in}}
\def\kcrit{k_{\rm crit}}
\def\Rcoll{R_{\rm coll}}

\def\Gc{{\rm G}}

\def\sec{\,{\rm s}}

\def\km{\,{\rm km}}
\def\pc{\,{\rm pc}}
\def\kpc{\,{\rm kpc}}
\def\Msol{\,M_\odot}
\def\Gyr{\,{\rm Gyr}}
\def\Myr{\,{\rm Myr}}
\def\pcspkm{{\,\rm pc\,s\,km^{-1}}}
\def\kmppcs{{\,\rm km\,s^{-1}\,pc^{-1}}}
\def\kmps{{\,\rm km\,s^{-1}}}

\let\theta=\vartheta
\let\phi=\varphi     

\input epsf

\catcode`@=11 

\ifshowmods
\input pstricks
\def\modins#1{{\bf #1}}
\def\moddel#1{{\lightgray #1}}
\else
\def\modins#1{#1}
\def\moddel#1{\relax}
\fi

\newtoks\refname
\let\referr@r=\relax 

\def\meqref #1{\global\advance\formcount by 1 \refname={#1}%
      \expandafter\xdef\csname 
      formref\the\refname\endcsname{\number\formcount}}
\def\eqref #1{\refname={#1}%
\expandafter\ifx\csname formref\the\refname\endcsname\relax\referr@r
\else\csname formref\the\refname\endcsname\fi}
\newcount\formcount
\formcount=0
\let\plaineqno=\eqno
\def\eqno#1{\meqref{#1} \plaineqno{(\number\formcount)}}

\def\makefigref #1{\global\advance\abbcount by 
      1 \refname={#1}%
      \expandafter\xdef\csname abbref\the\refname\endcsname{\number\abbcount}}
\def\figref #1{\refname={#1}%
\expandafter\ifx\csname abbref\the\refname\endcsname\relax\referr@r
\else\csname abbref\the\refname\endcsname\fi}
\newcount\abbcount
\abbcount=0

\catcode`@=\active 

\titlea{Introduction}

Since the work of Orr (1907) it has been known that certain 
perturbations in stable shearing flows show a transient 
exponential growth. This phenomenon was discussed extensively by 
Goldreich \& Lynden-Bell (1965) and Julian \& Toomre 
(1966) in the context of galactic dynamics, using hydrodynamics and statistical mechanics, respectively. Both 
papers examine the dynamics of a small patch within a 
larger disk and find strong transient 
growth of density perturbations \moddel{that grow} as they ``swing by'' from 
leading to trailing.  Toomre (1990) has argued that this amplification 
mechanism is the principal dynamical 
process responsible for spiral arms in disk galaxies.

While gas is known to play a major role in quasi-stationary density waves (Roberts 1969), there has been little research on the dynamical behavior of the interstellar medium (ISM)
in transient density waves.  It is this issue the present paper addresses.
The key question we want to answer is whether the short lifetime of a
swing-amplified perturbation still allows a noticeable response of the ISM. Additionally, we look for kinematical signatures that might allow a discrimination of quasi-stationary
versus transient density waves.

So far, the only study of swing amplification in a two-component
medium was
undertaken by Jog (1992) who extended the formalism of Goldreich \&
Tremaine (1978) to include two fluids with different stability numbers.
While these works used Lagrangian coordinates to solve the dynamical
equations, in this paper we employ a different approach using Eulerian
coordinates.  This enables us to consider a succession of
swing amplification events in a consistent way.

The outline of this paper is as follows:  In section~2 we describe our
model with a brief review of the results of Fuchs (1991) important to
this work and a
discussion of our
treatment of an ISM consisting of discrete clouds.  Section~3 presents the main
results.  In a first part, we describe the behavior of our model
under a single perturbation.  This behavior is interpreted in
the second part.  After a discussion of the model
with repeated perturbations, section~3 closes with an
investigation of
the effects of self gravity in the ISM.  In Section~4 we briefly
summarize the main conclusions we draw from this work.

\titlea{The Model}

Due to the inherently local nature of swing amplification no global disk model is
required. Instead, we use a sliding box scheme similar to the one used
by Toomre \& Kalnajs (1991).  The central idea of the sliding box scheme is to
examine the dynamics of a rectangular patch of a galactic disk that is
periodically repeated such that when, for example, a particle leaves the
patch at the rotationally leading border, it is re-fed at the trailing
one.  This resembles the periodic boundary conditions used in plasma and
solid state physics, with the complication that in galactic dynamics the
periodic continuations must share the general shear of the disk.  In
consequence, particles leaving the patch radially must be re-fed at the
opposite edge at positions varying with time, and the difference in
circular velocities between the inner and outer edges must be accounted
for, so as to keep the accelerations continuous during the re-feeding
process.

In such a patch pseudo-Cartesian coordinates $x=r-r_0$ and
$y=r_0(\theta-\Omega_0t)$ are introduced, where $r$, $\theta$ are
galactocentric polar coordinates, $r_0$ is the distance of the center of
the patch from the center of the galaxy, $\Omega_0$ is the angular
velocity at $r_0$, and $t$ denotes the time. After inserting these new
coordinates into the equations of motion derived from the Lagrangian
$L={1\over2}(\dot r^2+r^2\dot\theta^2)+\Psi(r,\theta)$ with the
potential $\Psi(r,\theta)$, we linearize the equations with respect to
$x$ and $y$. The linearization is valid under the assumption that (a) the
radial extent of the patch is small compared to the scales over which
the basic state of the galaxy varies significantly and (b) the
peculiar motions are much smaller than the circular
velocities.  Note that no assumption about the circumferential extent of the patch
is necessary.

\ifproofmaking\begfig{\epsfxsize=8.8cm
\epsfbox{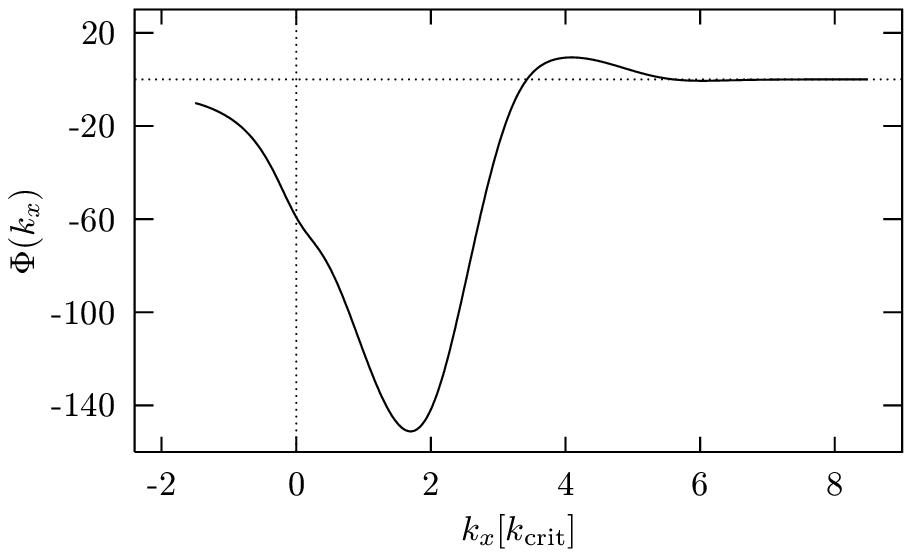}}cm\else\begfig 5.2cm\fi
\figure{\figref{thepot}}{Amplitude of the potential 
perturbation, given in arbitrary units.  The initial amplitude of the
potential perturbation at
$k_x=-1.5\kcrit$ and the minimum at about $k_x=1.6\kcrit$ correspond to  density perturbations 
of roughly $0.006\Sigma_0$ and $0.1\Sigma_0$ amplitude, respectively. Since the radial
wave number increases linearly with time, this also shows the temporal
development of the wave.   The interval of wave numbers shown here,
$k_x=-1.5\cdots8\,\kcrit$, corresponds to a time interval of $7\times 10^8$ years.}
\endfig

This leads to the epicyclic equations of motion
$$\meqref{moteq}\eqalignno{\ddot x&=2\Omega_0\dot
y+4\Omega_0A_0x+f_x&\rm(\eqref{moteq}a)\cr\ddot y&=f_y-2\dot
x\Omega_0,&\rm(\eqref{moteq}b)}$$ where $f_x$, $f_y$ are the
acceleration components due to the spiral field.  $A_0$ denotes
Oort's first constant.  The perturbation forces are calculated
according to the work of Fuchs (1991) where the Boltzmann and
Poisson equations were used to obtain the perturbation
potential in the form $$\Phi(x,y;t)=\Phi_0(t)\exp\bigl(\I\vec
k\cdot\vec r\bigr),\meqref{poteq} \eqno{(\eqref{poteq})}$$
where $\vec k=\bigl(k_x(t),k_y)$ is the wave vector of the
perturbation.  Since swing amplification is most effective in
the vicinity of $k_y=0.5\,\kcrit$, we use this choice
throughout the following.  In the case of free swing
amplification one has $k_x(t)=\kxin+2k_yA_0t$ for a sinusoidal
excitation with an initial radial wave number $\kxin$ and
assuming positive $k_y$.  While one has to expect that
excitations will not be sinusoidal in reality, the restriction
to a single Fourier component should not be critical due to
the linearity assumed in the derivation of Eq.~(\eqref{poteq})
and because components with wave numbers significantly
different from an optimal $\kxin$ will either be not amplified
strongly or damped out before reaching the zone of
amplification.

The integral  equation determining the amplitude function
$\Phi_0(t)$ derived by Fuchs (1991) is  solved numerically. 
Its solution is shown in  Fig.~\figref{thepot} for the
standard parameters of our  model, $$\eqalign{&\kcrit=2\pi/(10\kpc) 
\cr&\kxin=-1.5\,\kcrit \cr &Q^2=2}\quad\eqalign{&A_0=0.015\kmppcs\cr 
&B_0=-0.01\kmppcs\cr &\mu^{\rm in}/\Sigma_0=1/(50\pi),}$$ where 
$\kcrit=\kappa^2/2\pi\Gc\Sigma_0$ is Toomre's critical wave 
number, defined with the epicyclic frequency $\kappa$, the 
gravitational constant $\Gc$ and the unperturbed mass  surface
density $\Sigma_0$.  \modins{Our choice of $\kcrit$ gives
$\Sigma_0=58\Msol\pc^{-2}$, somewhat more than what is found for the Milky
Way in the solar neighbourhood (Kuiken \& Gilmore 1991).} 
$Q$ is the Toomre  stability parameter,
$\mu^{\rm in}$ is the amplitude of the  initial density
perturbation, and $B_0$ denotes Oort's second constant. 
\modins{With these parameters one has $\Omega_0=0.025\kmppcs$
with a locally slightly falling rotation curve. The only assumption
about the distance to the center of the disk is
$\Omega_0r_0\gg\sigma$.}
While the choice of $\kcrit$, $A_0$, $B_0$, and $Q$ is not
meant to accurately model the solar neighborhood, we regard it
as representative for the outer regions of giant
spirals.  \modins{Our results are generic in that
they do not qualitatively change under reasonable modifications
of these parameters.  For example, as long as $-B_0$
is comparable to $A_0$, it is unimportant whether the rotation
curve is flat, rising or falling.}

Equation~(\eqref{poteq}) describes 
sinusoidal waves of constant circumferential wavelength. 
Due to the shear of the disk, the radial wavenumber $k_x$ 
grows with time. In this way an initially leading 
wave ($\kxin<0$) is transformed into a trailing one and, as can be seen in Fig.~\figref{thepot}, 
strongly amplified while ``swinging by''. With increasing 
$k_x$ phase mixing leads to a damping of the perturbation
as the wavelength of $\Phi$ drops beyond the typical epicycle size. A single swing amplification event from excitation to decay has a duration of a few $10^8$ years.

In our model the ISM is approximated as a dissipative medium of
discrete  clouds. Their motions are governed by 
equations~(\eqref{moteq}) as long as they do not collide. 
Collision detection is performed on a grid of width $\Rcoll$, 
where we chose $\Rcoll=50\pc$. If there are two or more clouds
in one cell, we form pairs by a random process and let a pair
collide if the clouds that make it up are approaching each
other. The collisions  treat clouds as ``sticky'' spheres,
i.e.~their relative  velocities are multiplied by a factor $(1-f)$
after performing an elastic  two-body collision (Brahic 1977,
Schwarz 1981). \modins{In such a
system the cooling rate depends on the collision rate and
the inelasticy coefficient $f$.  To have a clear signature of
the collisonal nature of the cloudy medium, we have selected a high
collision rate (about $0.2\,{\rm step}^{-1}$) which in turn
requires a comparatively low $f$ of 0.2 to prevent
the system from cooling down too quickly.  For the implication
of this choice, see the comprehensive investigation by
Jungwiert \& Palous (1996).  However, as long as the total cooling
rate is kept constant, our model is much less
sensitive to the choice of $f$ than theirs.
With these
parameters, an unperturbed system of clouds with Gaussian
velocity distribution cools down
exponentially with a characteristic time of about $200\Myr$.}
\moddel{In our simulations  we chose an inelasticity
coefficient of $f=0.2$, which is rather low but allows for a
high collision rate without too much cooling.} Collisions are
checked for once every step, where one time step equals
$3\pcspkm$.

The sticky particles are initially homogeneously 
distributed with a density of $0.0003\pc^{-2}$ and 
have a Gaussian velocity distribution. We chose a radial 
velocity dispersion of 
$\sigma_u\approx4.2\km\sec^{-1}$. The velocity dispersion in the 
$y$-direction was set according to the epicyclic ratio
$(2\Omega_0/\kappa)$\modins{, resulting in a total initial velocity
dispersion of $\sigma=5\kmps$.  As long as $\sigma(t=0)$ is
low enough (say, $<10\kmps$), the behaviour of the model is not
strongly influenced by this choice.}

Velocity dispersions of this order are typical of giant 
molecular clouds (Clemens 1985). Our sticky particles, however,
are not  meant to mimic GMCs themselves. Their number is much
larger  than that of GMCs, they are stable, and of course
their dynamical behavior can at best be regarded as idealized
from real clouds. Technically, they  just implement a
dissipative medium. \moddel{The closest physical  counterpart might be
found in the constituent clouds of  GMCs. }\modins{From the structure
of the equations of motion, we do not need to make any assumptions
about the masses of the clouds.  Also, it is very hard to assign
radii to the model clouds from the sticky particles formalism. One
might use $\Rcoll/2$ as a gross measure of the collision cross section.}
The details of the
modeling do not seem to be critical for our  results, since
the dissipativity of the medium does not  influence its
dynamical evolution very significantly except through  cooling
(see below).  In this way, our results would be largely
unchanged if, for example, one increased $f$ and lowered
the initial density accordingly\modins{ to keep the collision
rate constant}.

Introducing two-particle interactions like collisions requires
 some care in strictly  two-dimensional calculations like the
present one. Rybicki  (1972) showed that in contrast to
three-dimensional systems in two-dimensional self-gravitating
sheets the relaxation time  is always of the order of the 
crossing time. Interactions ``softer'' than gravity are,
however, less affected by this limitation.  In fact, one can
choose $T_{\rm relax}/T_{\rm 
crossing}$ of the ISM freely in our model by changing $\Rcoll$
and the initial surface density since all interactions 
with impact parameters smaller than $\Rcoll$
are equivalent. Checks with  3-D-models showed no
significant deviations from the  evolution described below.

\ifproofmaking\begfig{\epsfxsize=8.8cm\epsfbox{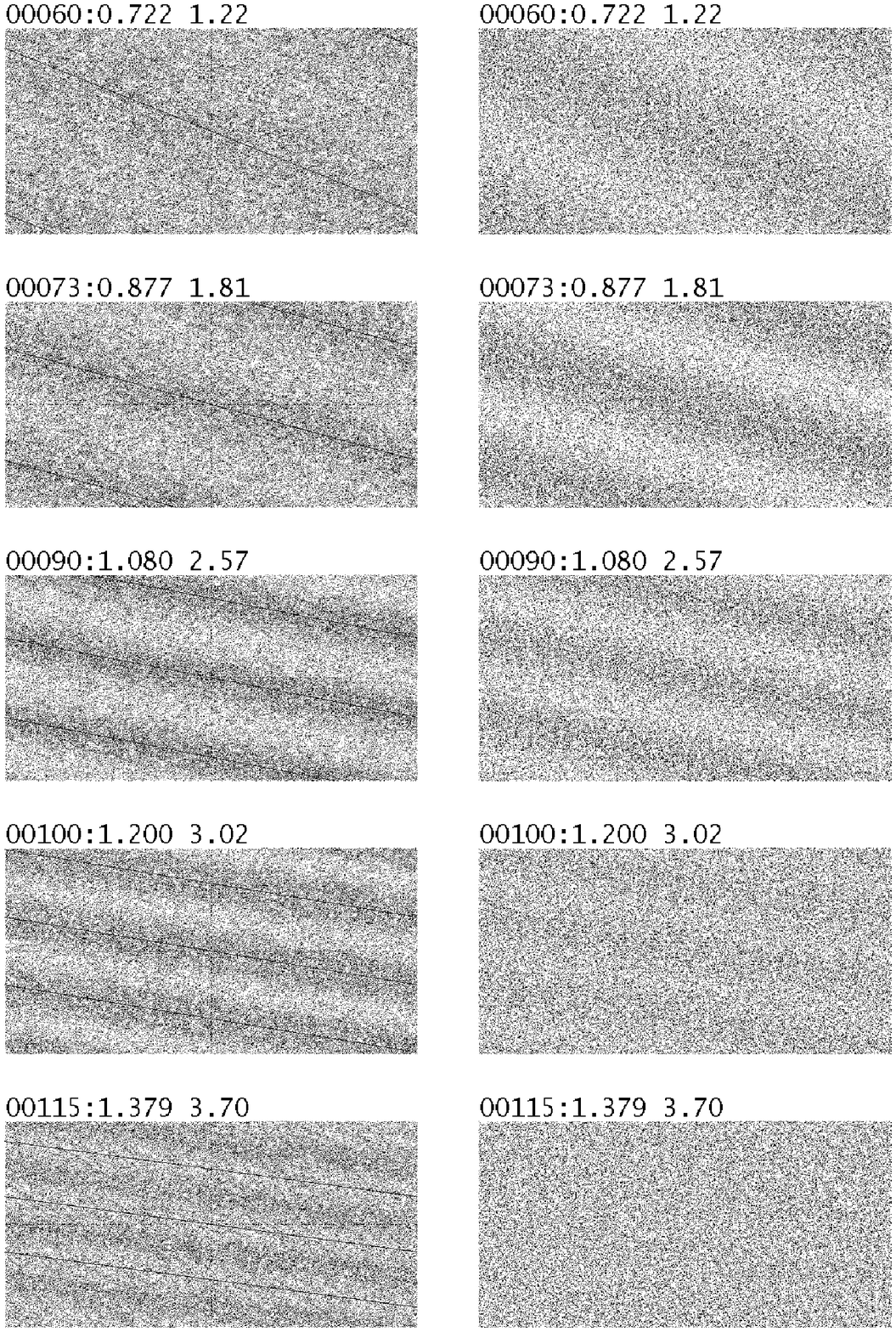}}cm\else\begfig13.5cm\fi
\figure{\figref{density}} {The distribution of gas clouds 
(left) and ``backbone'' stars (right) in a patch of dimensions 
$20\kpc\times10\kpc$\modins, giving 60000 clouds in the patch.  The radial direction is upwards and the galactic
rotation towards the right.
Above each panel the step 
number, the time since the excitation in units of galactic 
rotations (given for illustrative purposes only and computed for an assumed
$r_0=8\kpc$) and the $k_x$ of the perturbation in units of 
$\kcrit$ are indicated. In the frames on the left each dot 
represents a gas cloud and the straight lines mark phases 
zero and $\pm2\pi$ of the shearing wave. \moddel{On the right density plots of
the stellar mass density obtained from the Poisson equation are
shown.}
\modins{The frames on the right side are scatter plots of the
surface density $\Sigma(x,y)$ obtained from inserting 
(\eqref{poteq}) into the Poisson equation.}
To make the structures in the stellar disk visible, we enhanced the
contrast in the stellar density by a factor of four. }
\endfig

\titlea{Results}

\titleb{Evolution of a Single Density Perturbation}

In Fig.~\figref{density} we show the spatial distributions of
clouds and stars in a patch of a galactic disk constructed
in the way described above. We chose a rather 
large radial extent of the patch, $b_x=10\kpc$, in order to give a clear 
impression of what is going on. The physical relevance of 
the upper and lower edges of the displayed patch may 
therefore be doubtful, whereas in the region around $x=0$, 
located in the center of each frame, the linearization 
performed above is well justified.

\ifproofmaking\def\figarg{{\epsfxsize=12cm\epsfbox{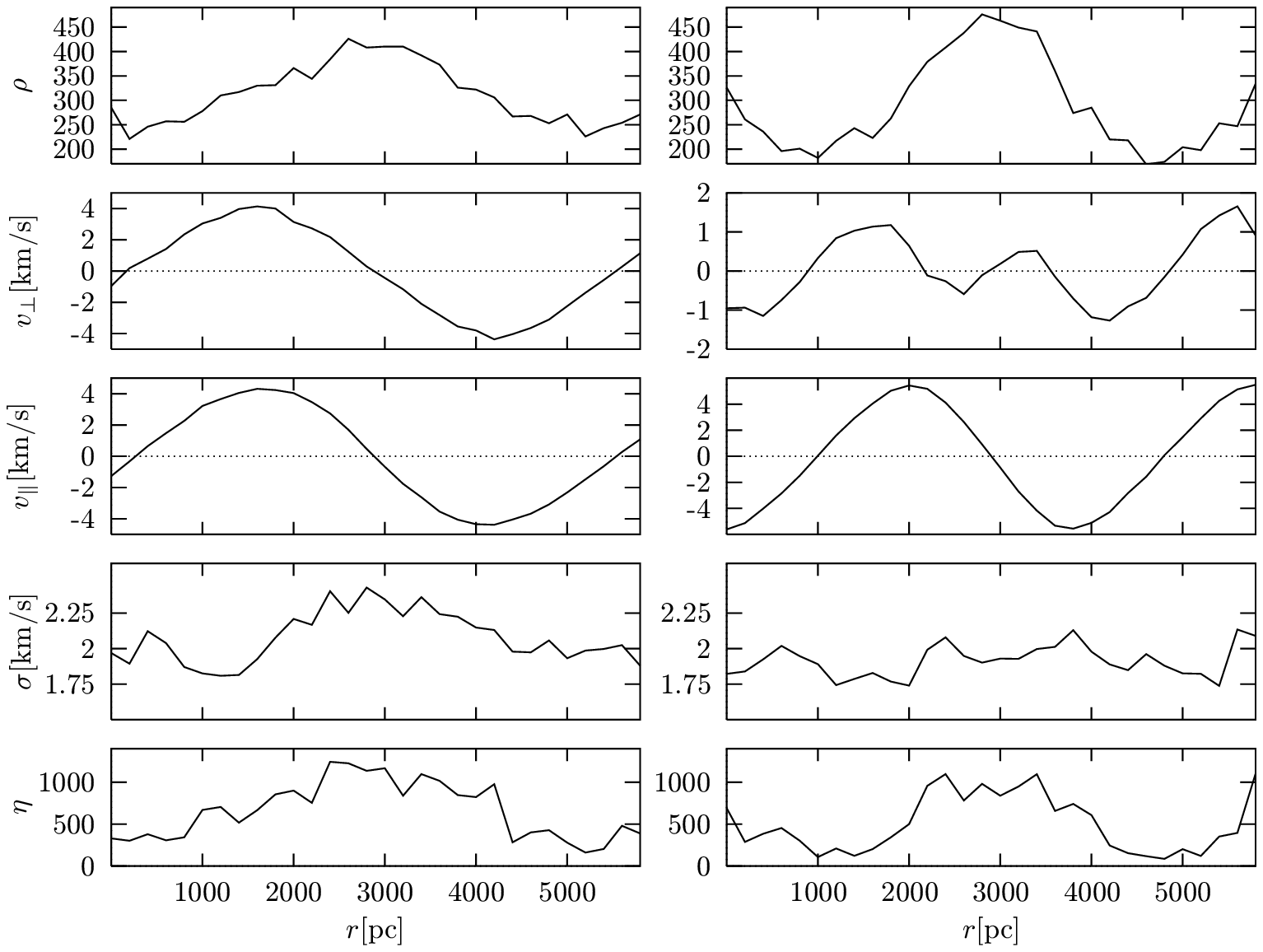}}cm 12cm}\else\def\figarg{8.8cm 12cm}\fi
\expandafter\begfigside\figarg
\figure{\figref{cuts}}{Profiles of the density $\rho$, the streaming
velocities perpendicular and parallel to the the wave crest after
subtracting the basic shear, $v_\perp$ and
$v_\parallel$, and the velocity dispersion, $\sigma$, in $\kmps$ across
a spiral arm, and the energy dissipated in collisions per unit time and
area, again in arbitrary units.  The profiles on the left have been
taken at step~73 and illustrate the state of the patch during a
formation phase of the spiral arms.  On the right we illustrate the
onset of the
dissolution phase with profiles taken at step~90.  These profiles were
obtained along lines of $6\kpc$
length perpendicular to the density wave crest. The wave crest is at
$r=3\kpc$.} \endfig

The first frame in the series shows the state of the 
patch about three quarters of a galactic rotation after the 
perturbation has been excited. The perturbation is already 
trailing and well developed in the galaxy's stellar ``backbone'' on 
the right.  The gas has responded to the potential perturbation and
the amplitude of the density perturbation in the gas is comparable to the one in the stars.

This changes about $30\Myr$ later, when the arm-interarm density contrast in the
stars has just passed its maximum of about 1.4:1.  As can be seen from Fig.~\figref{cuts} the density contrast in the gas is almost 2:1 at this stage.   The
flow pattern is characterized by an inflow towards the
wave crest and a tangential flow that transports clouds inwards on the
outer edge of the arm and outwards on the inner edge.  Judging from
the sinusoidal profiles, we still are in the linear regime.  These
profiles resemble analogous results from
simulations with discrete clouds in quasi-stationary density waves
(the results of Roberts (1992) are particularly well suited for
comparison). The main difference is that shearing density
waves are essentially at rest with respect to the disk material around them and thus
produce structures symmetric to the wave crest, whereas
stationary density waves outside the corotation zone have an upstream
and downstream
side and thus lead to asymmetric profiles.  

While the perturbation amplitude in the stars declines from
now on (cf.~Fig.~\figref{thepot}), the spiral arm in the gas
continues to grow for another $50\Myr$, as seen in the
third frame in Fig.~\figref{density}.  With this maximal
density contrast in the gas, the formation phase of the spiral
arm comes to an end.  The following dissolution phase is
marked by a gradual turnover from inflow to outflow with
respect to the wave crest.  In the velocity component
perpendicular to the wave crest, $v_\perp$, a double-wave
pattern evolves (cf.~Fig.~\figref{cuts}).

As can be seen in the fourth frame in Fig.~\figref{density},
the spiral arms widen\moddel{s} during the dissolution phase, and
\moddel{its}\modins{their}
edges sharpen.  This shift of activity towards the edges is 
discernible in the $\sigma$-profile shown in
Fig.~\figref{cuts}, right side, where one already has maxima
at the arm's edges. At this stage hardly any perturbation is
visible in the stellar disk. 

In the last frame shown in Fig.~\figref{density}, taken about
half an epicyclic period after the strongest expression of the
pattern, the gas clouds once more gather into a density
concentration, this time at phases $\pm(2j+1)\pi$, $j\in\bbbn$, of the
original density perturbation.  Apart from the shift in phase
and the smaller wavelength, profiles in $\rho$, $v_\perp$, and
$v_\parallel$ across this ``echo'' arm closely resemble those
given in Fig.~\figref{cuts}, left column.  However, it has no
visible counterpart in the stellar disk. Another half epicyclic
period later one again finds an echo, this time with
maxima at the positions of those of the original density perturbation. 
These echoes are only slowly damped out.

\ifproofmaking\def\figarg{{\epsfxsize=11cm\epsfbox{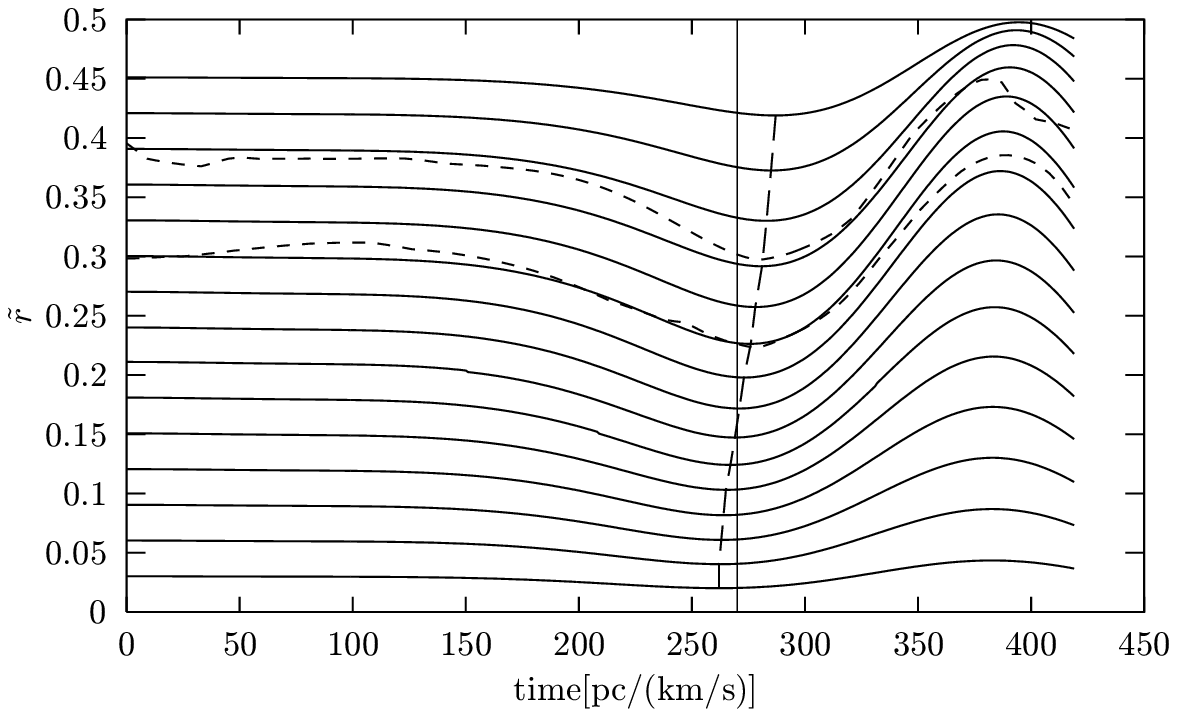}}cm 11cm}
\else\def\figarg{6.4cm 11cm}\fi
\expandafter\begfigside\figarg
\figure{\figref{key}}{Time evolution of crest distances.  Here we show
the distances of 15 collisionless particles starting with zero peculiar
velocity (solid lines) in units of the time-dependent wavelength of the
potential perturbation.  The long-dashed line connects the epochs
of the turnover of the particles' motion, and the
vertical line marks the time at which the profiles on the right hand
side of
Fig.~\figref{cuts} were taken.  The short-dashed lines indicate crest distances of two gas clouds from our reference
simulation for comparison.  Note that $\lambda(t)$
varies between $20\kpc$ and $2\kpc$ in the time interval shown in this diagram,
hence the amplitude of the oscillation is no measure for the sizes of the
epicycles.}\endfig

\titleb{Physical Interpretation}

To get an understanding of the dynamics of the patch, 
it is advantageous to introduce the crest distance
in units of the perturbation wave length, $\tilde r(t)=\vec
r\vec k/(2\pi)$. In $\tilde r$ unperturbed circular orbits
appear as parallels to the $x$-axis, and the acceleration due
to the perturbation is always parallel to the measuring rod. 
The drawback is that the length unit varies with time, so that
orbits of constant nonzero epicyclic amplitude appear to have
growing amplitude in $\tilde r$ when $k_x>0$.

The evolution of $\tilde r(t)$ for 15~particles moving in our
reference model without collisions is shown in
Fig.~\figref{key}.  The particles start on circular orbits and
are accelerated towards the wave crest for the first
$200\pcspkm$, indicated by a negative curvature in $\tilde
r(t)$.  At about the time when the stellar arm starts to
dissolve ($k_x\approx1.7\,\kcrit$ by Fig.~\figref{thepot},
corresponding to $t\approx210\pcspkm$), the curvature changes
\moddel{its} sign.  The net force on a particle must therefore be directed
away from the wave crest, although $\Phi_0(t)$ is still
negative. This is because the Coriolis force present in the
equations of motion (\eqref{moteq}) is now stronger than the
perturbation force.  While the latter decays, the Coriolis
force finally turns around the motion.  Since the relative
strengths of perturbation and Coriolis forces vary with the
crest distance, the epoch of this turnover
also depends on $\tilde r$.  This delay in turnover time causes the double wave
shown in Fig.~\figref{cuts} on the right side.

\meqref{radia}

\ifproofmaking\begfig{\epsfxsize=8.8cm\epsfbox{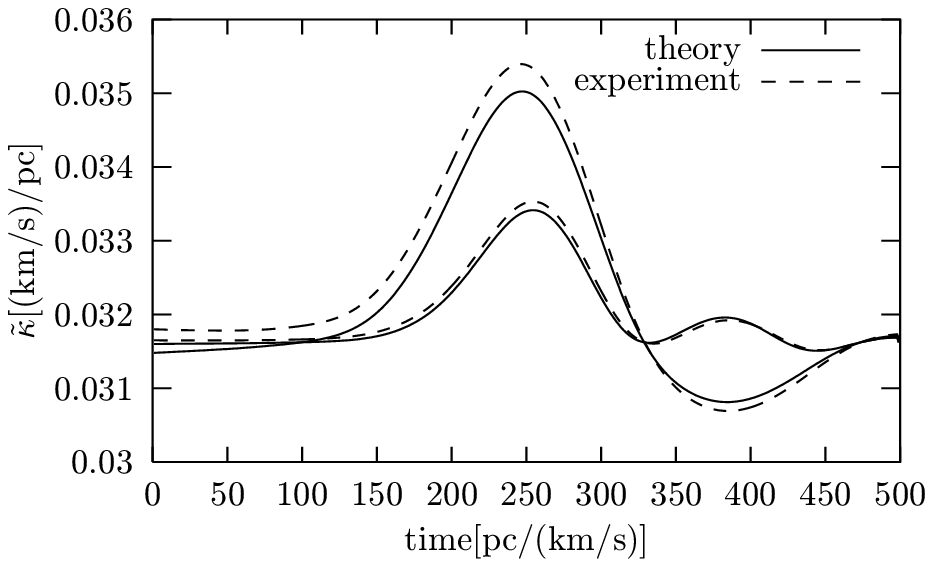}}cm\else
\begfig5.2cm\fi
\figure{\figref{chepi}}{
Perturbed epicyclic frequencies $\tilde\kappa$ for two distances from
the wave crest as a function of time.  The solid lines are computed
from Eq.~(\eqref{radia}).  The dashed lines are averages for the
epicyclic frequencies over the numerically integrated orbits
of 250 particles distributed evenly
on an epicycle with $u=10\kmps$.  For the upper curve, the guiding
center of the epicycle lies at $x=200\kpc$, for the lower curve
at $x=1500\kpc$.
}
\endfig

While it is possible to gain a qualitative understanding of the
variation of turnover times in terms of the ratio of perturbation versus
Coriolis force, the problem is better understood in terms of 
a perturbation of the epicyclic frequency, i.e.~by letting the angular
frequency of a supposed periodic motion in a shearing reference frame vary with the distance
to the wave crest.  For a
simplified version of the present potential, we derive $${\tilde\kappa^2(r)\over\kappa^2}\approx\left(1-{ k\alpha\over\kappa^2}\cos(k
r)\right)\eqno{(\eqref{radia})}$$ in the appendix for the perturbed epicyclic frequency $\tilde\kappa$
where $\alpha$ is the amplitude of the perturbation\modins{---which is 
negative during the formation phase---}, $k$ is its wave
number and $r$ denotes the distance from the wave crest.  Thus, the
double wave is a consequence of $\tilde\kappa$ increasing towards the wave
crest.  As can be seen from Fig.~\figref{chepi}, the rough estimate
Eq.~(\eqref{radia}) already is quite a good
approximation to the more complex situation in our simulation. 

Fig.~\figref{key} also shows how the dissolution phase turns over into
the formation of the echo arm as the clouds proceed on their epicycles.
After $t=300\Myr$ ($k_x=3\,\kcrit$, cf.~Fig.~\figref{thepot}) the
potential perturbation is practically zero, and therefore the echoes
simply are kinematical spiral arms (Kalnajs 1971). They do not appear in the stellar disk because of the much higher
velocity dispersion of the stars. For our choice of $k_y=0.5\,\kcrit$,
the size of the stellar epicycles gets comparable to the perturbation
wavelength shortly after the maximal amplification, and phase mixing
leads to an exponential damping.  Thus the only sign of kinematical
spiral arms in the backbone stars is a small overshoot of $\Phi_0$
around $k_x=4\,\kcrit$ (cf.~Fig.~\figref{thepot}).

In contrast, the echo arm in the gas has about the
amplitude of the primary spiral arm. This is mainly because of
the smaller velocity dispersion of the gas.  When phase mixing
becomes important in the stars, the wave length of the
perturbation is still much larger than the typical size of an
epicycle of a cloud.  Only at a wavelength of about $500\pc$ ($t>900\Myr$) does
phase mixing dominate the damping of the perturbation in the
gas.  Until then, the perturbation is mainly damped by
collisions that are not very efficient in damping away
larger-scale structures.
Thus, even when radial variations of
$\kappa$ weaken the kinematical spiral arm, at least the
first echo can be  expected to be reasonably strong.  One
might speculate  that echoes of this kind have some relation
to ``gaseous interarms'', spiral arms in the ISM without
corresponding activity in the stellar  disk (e.g.,
Block et al (1994), although their observations only cover
 spirals with
much higher density amplitudes than we consider here).

The discussion of the dynamical evolution of the patch given so
far did not make any reference to the dissipativity of the
cloudy medium.   In fact, turning collisions off does not
strongly change the evolution of the density perturbation in
the ISM.  The impact of the dissipativity on the dynamics is
small because the
streaming motions induced by the perturbation are quite smooth,
so that the
dissipativity of the medium mainly appears as viscosity that
can be disregarded on short time scales. On longer time scales,
however, the dissipativity of the cloudy medium keeps its
velocity dispersion low, whereas a single swing amplification
event roughly doubles the temperature of a stellar system with
$\sigma(t=0)=5\kmps$.

Spiral arms are traced by young stars and H{\sc II}-regions.
Although the question has been under some debate (Elmegreen
\& Elmegreen 1986), it appears that density waves do modulate
either the initial mass function or the star formation rate 
nonlinearly (Cepa \& Beckman 1990).  Kwan \& Valdes (1983) have
argued that massive star formation is triggered by collisions
between diffuse molecular clouds.  Following this line one
might expect the energy dissipated in collisions per unit time
and area, $\eta$, to give a coarse measure for the efficiency
of massive star formation.  \modins{Though this is a na\"\i ve
view of how star formation proceeds in molecular clouds, the
compression induced  by the development of spiral arms
certainly plays a role in triggering the collapse of molecular
clouds and hence star formation.} 
As can be seen from
Fig.~\figref{cuts} we find contrasts in $\eta$ of up to $1:10$
from arm to inter-arm regions since $\eta$ is larger in the arms.

\titleb{Consecutive Swing Amplification Events}

Certainly the processes investigated here will not take place only once
for each galaxy; instead, we expect excitations to recur in a more or
less stochastic manner. To examine the behavior of the patch under
repeated perturbations, we need to make some assumptions regarding the
frequency and the amplitude spectrum of the amplification seeds. In our
model, we are rather limited in the first of these choices because the
crests of the density waves must always connect the centers of adjacent patches
to keep the perturbation potential continuous across patch borders.
Consequently, the swing amplification events must follow one another
with a period of $b_y/(2A_0b_x)$ (or a multiple thereof), where the circumferential extent of
the patch is already fixed to $b_y=2\pi/k_y$ by the continuity condition across
circumferentially adjacent patches.  

In contrast, we expect swing amplification events to follow one
another in a more or less random fashion.  This should be the
case for seeds provided by Poisson noise or massive perturbers
of the disk (Toomre 1990).  Even recurrent swing amplification
due to mode coupling as studied by Fuchs (1991) will have a
stochastic nature.  Thus strictly periodic excitations are
probably grossly unphysical.  However, as neither the global
kinematics nor the overall morphology of the patch strongly
depends on the excitation period as long as one avoids
resonances with the epicyclic period, one can still gain some
insight in the long\moddel{er-} term evolution of the patch.   For
simplicity, we use the same amplitude for all seeds.

The general appearance of the patch subjected to such repeated
perturbations qualitatively resembles the sequence shown in
Fig.~\figref{density} for a single swing amplification event. 
However, the spiral arms are more ragged with crossings and
bridges, because the potential perturbations act on a disk
that is already structured both in density and velocity.
\modins{For excitation periods used here, only one swing amplification
event causes noticable perturbation forces at any moment.  This
corresponds to the fact that the lifetime of any perturbation will be
much shorter in the backbone stars with their high velocity
dispersion than in the cool cloudy medium.  One consequence of this is that the 
spiral arm in the backbone stars will be much smoother than in the
cloudy medium with no crossings and bridges appearing.  Again, this
invites comparison with the observations of Block et al (1994), who
see much smoother arm shapes in the infrared than in the optical.}

In longer-term simulations the gas reaches a quasi-equi\-lib\-r\-ium
state after the first few excitations. Then,
the global velocity dispersion including streaming motions
oscillates by about $1\kmps$ around a mean value of about
$\sigma\approx 6\kmps$ for our parameters.  This mean value 
increases slightly with decreasing $\tau_e$ over a range of
excitation periods from $150\pcspkm$ to $400\pcspkm$
(excluding resonances with $\kappa$). \modins{This equilibrium is
remarkably independent of the collision parameters.  Going from a mean
collision rate of $0.14\,{\rm step}^{-1}$ to $0.3\,{\rm step}^{-1}$
one finds the temporal mean of the velocity dispersion decreases by less than
$0.25\kmps$, where most of the difference is due to the systems with
high collision rates cooling down more quickly from a peak velocity
dispersion of about $6.5\kmps$ that is almost independent of the 
collision rate.
The reason for this behaviour is that 
in the later evolution $\sigma$ is dominated by the
thermalization of streaming motions.  These are to 
zeroeth order independent of the velocity dispersion before the
swing amplification event.  Thus, cool systems are
heated to the about the same temperature as warmer ones. 
As long as the system is able to cool down between two swing
amplification events, and provided
that the streaming motions are
not destroyed by too frequent collisions, the sole difference between
the cooler and warmer systems lies in the the amplitude of the
temperature oscillations.  Only when collisions become
unimportant---at collision rates of, say, $0.05\,{\rm step}^{-1}$---,
does this
reasoning break down and the mean velocity dispersion rises to values
of the order of $10\kmps$.  The velocity dispersion of a 
collection of collisionless test particles in the perturbed disk rises
to about $20\kmps$ within $3\Gyr$, which is still much less than
the $45\kmps$ of a stellar disk with the values of $Q$ and $\Sigma_0$ we
assume.}

\ifproofmaking\begfig{\epsfxsize=8.8cm\epsfbox{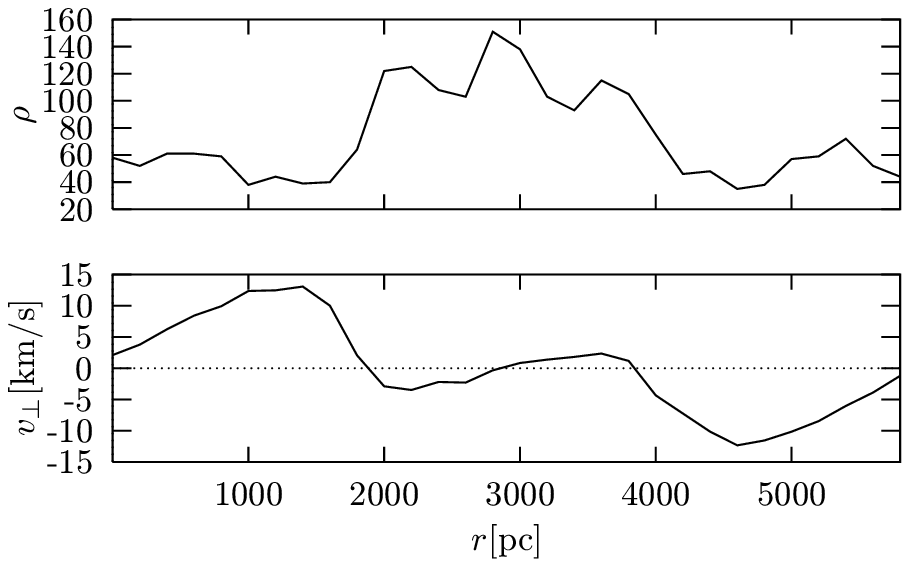}}\else\begfig5.2cm\fi
\figure{\figref{consqt}}{Profiles of $\rho$ and $v_\perp$
analogous to the ones shown in Fig.~\figref{cuts}, this time taken
after 1.5~Gyr of evolution under consecutive swing amplification
events.}
\endfig

The profiles in Fig.~\figref{consqt} show that by constructive interference between two swing amplification events a
pre-structured ISM can show a much more vigorous response to a
perturbation than the homogeneous patch examined above. Of course, destructive interference also occurs.  Finding profiles like the one shown in Fig.~\figref{consqt} should be quite within reach of current observational
techniques.  

One also finds that the double-wave pattern can persist
for up to $1/3$ of the lifetime of a spiral arm, depending on how the
excitations follow one another.

\titleb{Self gravity}

Jog (1992) predicts from her two-fluid\modins{s} model that the oscillation frequency for spiral arms in
the gas is much lower than for the stars from a
self-consistent two-fluid model.  This is not the case in our cloudy
medium.  In fact, kinematical spiral arms in an ISM consisting of
massless clouds will oscillate with the epicyclic frequency.  The
question arising naturally is whether the oscillation frequencies in
gas and stars are so similar because of the neglect of self gravity
in the ISM.

The investigation of the effects of self gravity could be implemented 
by a straightforward $N$-body code for our discrete clouds. However, 
considering Fig.~\figref{density} one notices that the
density evolution in the gas has the form of a shearing wave that
closely follows the evolution of the stellar perturbation in
wavelength and inclination.  Thus
the additional perturbation potential arising from the gas clouds can
be taken into account by assuming that the surface density of the gas
sheet is modulated like $\mu_{\rm gas}\propto\exp(i\vec k\cdot\vec r)$ with
$\vec k=(\kxin+2A_0k_yt,k_y)$.

Assuming a linear response the Poisson equation 
yields the additional potential perturbation arising from self gravity
$$\Phi_{\rm self}(x,y;t)=K{|\vec k^{\rm in}|\over |\vec
k(t)|}{\epsilon\Sigma_0\over \mu^{\rm in}}\Phi_0(\kxin)\cos(\vec
k\cdot\vec
r),\meqref{selfcon}\eqno{(\eqref{selfcon})}$$ where $\vec k^{\rm in}$ denotes the wave vector of the initial
perturbation, $\epsilon=\mu_{\rm gas}/\mu_{\rm stars}$ and $K$ is the
Fourier amplitude of the perturbation in the gas on $\cos(\vec k\vec
r)$. In practice, $K$ is
determined by a least-squares fit of $K\cos(\vec k\cdot\vec r)+\bar\mu$ to a
density cut perpendicular to the wave crest.

The assumption of a linear response is certainly valid in the early
stages of the evolution, while its justification is somewhat doubtful
for the later stages of the dissolution phase when the edges of the arm
get very sharp. Still, one may expect results that allow a qualitative
judgment on the importance of self gravity.

\ifproofmaking\begfig{\epsfxsize=8.8cm\epsfbox{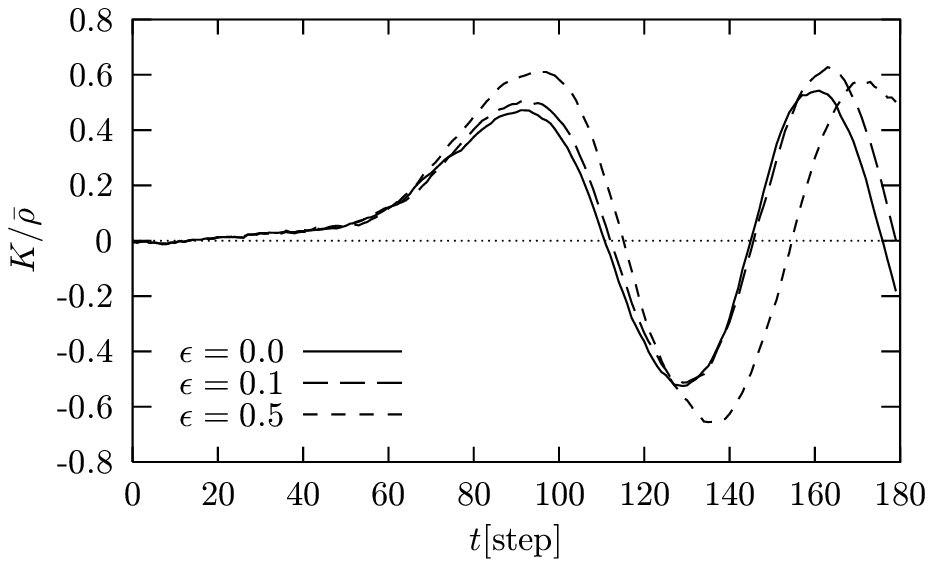}}cm\else\begfig5.2cm\fi
\figure{\figref{figjog}}{Time evolution of the density contrast of the
spiral arm for three values of the gas-to-stars mass ratio. For a low
gas mass fraction, the effect of self gravity is negligible.}
\endfig

Figure \figref{figjog} shows that neglecting self gravity is well
justified in our model for gas mass fractions similar to those found in
the solar neighborhood; the oscillation frequency of the arm
in the gas does not change much for $\epsilon=0.1$.  
For late-type spirals with a high gas content,
however, self gravity alters the behavior significantly and our results
should be applied with care. In particular, looking at profiles
analogous to Fig.~\figref{cuts} one finds that a higher gas mass fraction
will lead to softer edges and a less pronounced double wave during the
dissolution phase.

\modins{Of course, this scheme, being designed to assess the
consequences of neglecting self gravity on the large scale evolution
in our two dimensional sheet, cannot describe many effects certainly
important to disk evolution.  In particular, Toomre (1990) pointed
out that an $N$-body simulation in a patch similar to ours  rapidly
develops GMC-like complexes that might act as seeds for swing
amplification.  One can expect these to occur as well in the cloudy
medium investigated here if it were fully self gravitating.  
However, an examination of these processes,
possibly leading to self-triggered swing amplification with  
gas coupling back on the stars, is beyond the scope of this paper.}

\titlea{Conclusion}

In this work we used a model consisting of a stellar continuum and gas clouds
behaving like sticky particles to examine the behavior of gas in transient 
density waves.  In contrast to the prior work of Jog (1992) that
employed a gas-dynamical approach we find that the evolution of the
density perturbation in the gas
remains closely coupled to the one in the stars even when self gravity of
the ISM is taken into account.

In a first phase, the ISM responds to the growing potential
perturbation by flowing towards the wave crest.  When the
potential perturbation already decreases, the amplitude in the
gas continues to grow until the dissolution of the gaseous arm
begins due to Coriolis forces.  In this process, a variation of a
cloud's epicyclic frequency with its distance from the wave
crest leads to a double wave in the velocity perpendicular
to the arm, with particles near the wave crest already flowing
outward while those further out are still moving in.  We
regard this as a strong signature for transient density waves
that might be within reach of current observational
techniques. Finally, the gaseous arms continue to evolve as 
kinematical spiral arms until they are finally damped out by phase
mixing.

For the longer-term evolution we propose that swing
amplification events \modins{follow one another} in a more or less random fashion.  We find that
for a wide range of excitation frequencies spiral arms continue to form
and dissolve.  By interference between density waves a ragged morphology
results.  The dissipativity of the gas---unimportant for the short-term
dynamics---keeps the velocity dispersion of the clouds low and thus
their responsiveness high.

Given the stochastic nature of the processes described here, one
cannot predict the appearance of a galaxy based on a knowledge of its
basic state using the methods applied in this work.  However, since our results are quite insensitive to the
choice of parameters of the underlying model, we regard them as
fairly generic for transient density waves.  Thus, finding a double wave in radial velocity in the line of
sight across a spiral arm for some component of low velocity dispersion
(e.g., gas or OB-associations) would make a strong point for the
existence of transient spiral arms.

\acknow{I gratefully thank B.~Fuchs and R.~Wielen for many helpful
discussions and encouragement during the course of this work.
The comments of the referee M.~Gerin helped to improve the presentation 
of the model and the results.}

\begref{References}

\ref Bertin G., Lin C.C., Lowe S.A., Thurstans R.P., 1989, ApJ 338,
78

\ref Block D.L., Bertin G, Stockton A, Grosb\o l P, 
Moorwood A.F.M., Peletier R.F., 1994, A\&A 288, 
365

\ref Bogoliubov N.N., Mitropolski Y.A. 1961: Asymptotic Methods in the
Theory of Non-Linear Oscillations, Gordon\&Breach, New York

\ref Brahic A., 1977, A\&A 54, 895

\ref Cepa J., Beckman, J.E., 1990, ApJ 349, 497

\ref Clemens D.P., 1985, ApJ 295, 422

\ref Elmegreen, B.G., Elmegreen, M.E., 1986, ApJ 311, 554

\ref Fuchs B., 1991, In: Sundelius B. (Ed.), Dynamics of 
Disk Galaxies, G\"oteborg University, G\"oteborg, p.~359

\ref Goldreich P., Lynden-Bell D., 1965, MNRAS 130, 125

\ref Goldreich P., Tremaine S., 1978, ApJ 222, 850

\ref Jog C.S., 1992: ApJ 390, 378

\ref Julian W.H., Toomre A., 1966, ApJ 146, 810

\ref Jungwiert B., Palous J., 1996, A\&A 311, 397

\ref Kalnajs A., 1971, ApJ 166, 275

\ref Kuijken K., Gilmore G., 1991, ApJ 367, L9

\ref Kwan J., Valdes F., 1983, ApJ 271, 604

\ref Orr  W., 1907, Proc.~Roy.~Irish Acad.~{\bf A27}, 69

\ref Roberts W.W., Jr., 1969, ApJ 158, 123

\ref Roberts W.W.~Jr., 1992, In: Dermott S.F., Hunter J.H., 
Wilson R.E.~(Eds.): Astrophysical Disks, New York, p.~93

\ref Rybicki G., 1972, In: Lecar M. (Ed.), Gravitational $N$-Body Problem 
(Proceedings of the tenth IAU-Colloquium), Reidel, Dordrecht, p.~22

\ref Schwarz M.P., 1981, ApJ 247, 77

\ref Toomre A.~1964: ApJ 139, 1218

\ref Toomre A., 1990, In: Wielen R.~(Ed.), Dynamics and 
Interactions of Galaxies, Springer, Berlin, Heidelberg, New York, 
p.~292

\ref Toomre A., Kalnajs A.~1991, in Sundelius B. (Eds.), 
Dynamics of Disk Galaxies, G\"oteborg University, G\"oteborg, p.~339

\endref

\appendix{}

In the special case of a weak and purely radial perturbation,
$(f_x,f_y)=(\alpha\sin(kx),0)$, it is easy to show the dependence of the
local oscillation frequency on the position of the particle within the
density wave.  While this is quite different from the potential
arising from a shearing density wave, one can readily verify by
numerical integration of the orbits
that the oscillation frequency of mass points in a potential of the
type of Eq.~(\eqref{poteq}) roughly follows the behaviour predicted here with
minor deviations resulting from the time-dependence of the
wavelength of the perturbation (Fig.~\figref{chepi}).

The assumption of a purely radial perturbation allows a closed
integration of 
Eq.~(\eqref{moteq}b). Inserting the result into
Eq.~(\eqref{moteq}a) yields $$\ddot x=2\Omega
v_0-\kappa^2x+\alpha\sin(kx),\meqref{tgs}\eqno{(\rm A1)}$$ where
$v_0$ is an integration constant.  To the author's knowledge, a closed
solution of this differential equation does not exist.  We therefore
chose to investigate its solutions using the method of harmonic
balance (e.g., Bogoliubov \&
Mitropolski 1961).  We assume a quasi-harmonic solution of the
form $$x(t)=x_0+x_1\sin(\tilde\kappa t),$$ where
$\tilde\kappa=\tilde\kappa(x_0,x_1)$.  Inserting this into
Eq.~(A1) and Fourier analyzing the nonlinear term leads to
$$\eqalign{-\tilde\kappa^2x_1\sin(\tilde\kappa
t)&=\kappa^2x_0-2\Omega_0v_0+\alpha\sin(kx_x)J_0(kx_1)\cr
&+\bigl(\kappa^2-{2\alpha\over
x_1}\cos(kx_0)J_1(kx_1)\bigr)x_1\sin(\tilde\kappa t)\cr&+\hbox{higher
Fourier components,}}\eqno{\rm(A2)}$$ where
$J_0$ and $J_1$ denote the Bessel functions of order~0 and 1,
respectively.  By comparing the `leading' terms proportional to $\sin(\tilde\kappa t)$ one obtains
$$\tilde\kappa^2=\kappa^2\,\left(1-{2\alpha\over\kappa^2
x_1}J_1(kx_1)\cos(kx_0)\right).\eqno{\rm(A3)}$$ 

A straightforward application of (A3) to the perturbation used in our
simulations---justified {\it a posteriori} by Fig.~\figref{chepi}---can be done by
identifying $x_0$ with the distance of the guiding center of a
particle from the wave crest.  Since the epicycle sizes are small
compared to the wavelegth $2\pi/k$ of the perturbation, one may
furthermore approximate $J_1(kx_1)/x_1\approx k/2$.  This leads to
Eq.~(\eqref{radia}).  These approximations obviously break down for epicycle sizes of the order of a typical stellar epicycle size 
where the amplitude of the oscillation in $\tilde\kappa$ has to be zero.

\bye